\documentclass[a4paper,11pt]{article}
\pdfoutput=1 % if your are submitting a pdflatex (i.e. if you have
\usepackage{jinstpub} % for details on the use of the package, please
\usepackage{listings}
\title{\boldmath The Upgrade I of LHCb VELO - towards an intelligent monitoring platform}

\author[a]{P. Kopciewicz,} \author[a]{T. Szumlak,}
\author[a]{M. Majewski,} \author[b]{K. Akiba,}
\author[c]{O. Augusto,} \author[d]{J. Back,}
\author[e]{D. S. Bobulska,} \author[f]{G. Bogdanova,}
\author[g]{S. Borghi,} \author[h]{T. Bowcock,}
\author[c]{J. Buytaert,} \author[i]{E. Lemos Cid,}
\author[c]{V. Coco,} \author[c]{P. Collins,}
\author[b]{E. Dall'Occo	,} \author[c]{K. de Bruyn,}
\author[g]{S. de Capua,} \author[h]{F. Dettori,}
\author[h]{K. Dreimanis,} \author[g]{D. Dutta,}
\author[e]{L. Eklund,} \author[c]{T. Evans,}
\author[c]{M. Ferro-Luzzi}
\author[c]{W. Funk,} \author[j]{L. Meyer Garcia,}
\author[i]{O. Boente García,} \author[g]{M. Gersabeck,}
\author[d]{T. Gershon,} \author[b]{C. Sanchez Graz,}
\author[h]{K. Hennessy,} \author[b]{W. Hulsbergen,}
\author[h]{D. Hutchcroft,} \author[b]{D. Hynds,}
\author[k]{P. Jalocha,} \author[k]{M. John,}
\author[k]{N. Jurik,} \author[b]{I. Kostiuk,}
\author[d]{T. Latham,} \author[f]{A. Leflat,}
\author[h]{V. Franco Lima,} \author[j]{F. Marinho,}
\author[j]{L. H. Mendes,} \author[b]{M. Merk,}
\author[d]{A. Morris,} \author[g]{D. Murray,}
\author[e]{S. Naik,} \author[j]{I. Nasteva,}
\author[a]{A. Oblakowska-Mucha,} \author[j]{J. Otalora,}
\author[g]{C. Parkes,} \author[i]{B. G. Plana,}
\author[i]{A. Fernández Prieto,} \author[a]{B. Rachwal,}
\author[i]{P. Vázquez Regueiro,} \author[h]{K. Rinnert,}
\author[j]{G. Rodrigues,} \author[k]{L. Scantlebury-Smead,}
\author[e]{M. Schiller,} \author[c]{H. Schindler,}
\author[h]{T. Shears,} \author[b]{A. Snoch,}
\author[g]{P. Švihra,} \author[i]{A. Gallas Torreira,}
\author[b]{M. Van Beuzekom,} \author[l]{J. Velthuis,}
\author[f]{V. Volkov,} \author[g]{M. Williams}

\affiliation[a]{AGH University of Science and Technology,\\Krakow, Poland}
\affiliation[b]{Nikhef National Institute for Subatomic Physics,\\ Amsterdam, The Netherlands}
\affiliation[c]{European Organization for Nuclear Research (CERN),\\ Geneva, Switzerland}
\affiliation[d]{Department of Physics, University of Warwick,\\ Coventry, United Kingdom}
\affiliation[e]{School of Physics and Astronomy, University of Glasgow,\\ Glasgow, United Kingdom}
\affiliation[f]{Institute of Nuclear Physics, Moscow State University (SINP MSU),\\ Moscow, Russia}
\affiliation[g]{School of Physics and Astronomy, University of Manchester,\\ Manchester, United Kingdom}
\affiliation[h]{Oliver Lodge Laboratory, University of Liverpool,\\ Liverpool, United Kingdom}
\affiliation[i]{University of Santiago de Compostela,\\ Santiago de Compostela, Spain }
\affiliation[j]{Universidade Federal do Rio de Janeiro (UFRJ),\\ Rio de Janeiro, Brazil}
\affiliation[k]{Department of Physics, University of Oxford,\\ Oxford, United Kingdom}
\affiliation[l]{H.H. Wills Physics Laboratory, University of Bristol,\\ Bristol, United Kingdom}
\abstract{The Large Hadron Collider beauty (LHCb) detector is designed to detect decays of b- and c- hadrons for the study of CP violation and rare decays. At the end of the LHC Run 2, many of the LHCb measurements remained statistically dominated. In order to increase the trigger yield
for purely hadronic channels, the hardware trigger will be removed, and the detector will be read out
at 40 MHz. This, in combination with the five-fold increase in luminosity, requires radical changes to
LHCb's electronics, and, in some cases, the replacement of entire sub-detectors with state-of-the-art
detector technologies. The Vertex Locator (VELO) surrounding the interaction region is
used to reconstruct the collision points (primary vertices) and decay vertices of long-lived particles (secondary vertices). The upgraded VELO will be composed of 52 modules placed along the beam axis divided into two retractable halves. The modules will each be equipped with 4 silicon hybrid pixel tiles, each read out by 3 VeloPix ASICs. The total output data rate anticipated for the whole detector will be around 1.6 Tbit/s. The highest occupancy ASICs will have pixel hit rates of approximately 900 Mhit/s, with the corresponding output data rate of 15 Gbit/s. The LHCb upgrade detector will be the first detector to read out at the full LHC rate of 40 MHz. The VELO upgrade will utilize the latest detector technologies to read out at this rate while maintaining the required radiation-hard profile and minimizing the detector material.
}


\proceeding {15th Topical Seminar on Innovative Particle and Radiation Detectors\\
  14-17 October 2019\\
  Siena, Italy}

\begin{document}
\maketitle
\flushbottom

\section{Introduction}

LHCb (Large Hadron Collider beauty) \cite{lhcb} is a single-arm forward detector, dedicated to the study of heavy quark flavor sector, especially the charge-parity violation (CPV) phenomena, at the Large Hadron Collider (LHC). Its coverage in pseudorapidity, $\eta$, is complementary to the general-purpose experiments currently operating at the LHC. After the first phase of operation, from 2010 until 2018 called Run 1 and Run 2, LHCb is currently undergoing a deep upgrade of many crucial sections of the experimental setup \cite{intent}. The primary purpose of this upgrade is to take advantage of the increased number of concurrent collisions per beam crossing (pile-up), which will grow from 1.4 to about 5 in Run 3, to access rarer processes or statistically limited measurements.
These new operating conditions and an increase in fast hadron fluence motivated the whole upgrade program. Also, in order to cope with more busy events, a new DAQ system, featuring flexible full-software trigger capable of full detector readout at the LHC machine clock, had to be designed. In addition, to increase the number of stored events, it is planned that the raw data will be discarded.

The new vertex detector (VELO), based on silicon pixel sensors, is a vital part of the LHCb upgrade project. It will be readout at 40 MHz and operate at luminosities up to $2\times10^{33}$ cm$^{-2}$s$^{-1}$. The active pixels will be as close as 3.5 mm from the proton beams. The data rates will reach 1.6 Tbit/s, and the maximum 1-MeV neutron-equivalent fluence will reach $8 \times 10^{15}~1$ cm$^{-2}$ at the tip of the innermost sectors. A complete description of the new VELO detector can be found in Ref. \cite{Collaboration:1624070}.

The remainder of this paper is divided as follows: in Section 2, a detailed description of the VELO sensors and its calibration is given. In Section 3, an introduction to the new VELO software platform is presented and followed by a discussion of selected monitoring tasks. The paper is finished with a summary.

\section{The VeloPix sensor and calibration procedure}

\begin{figure}[h!]
    \centering
 \includegraphics[scale=0.33]{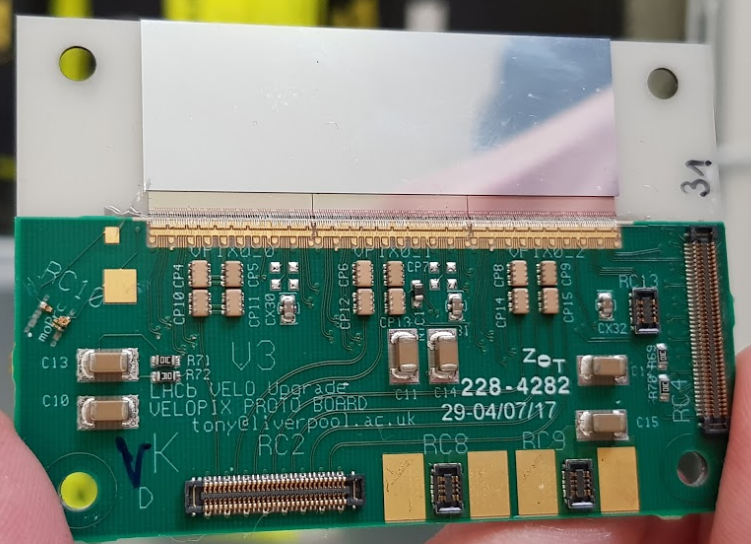} 
    \caption{One of the VeloPix sensor triplet prototypes.}%
    \label{fig:example}%
\end{figure}

The pixel sensor read-out chip designed for the upgraded VELO is a VeloPix ASIC \cite{Poikela_2017}. Each ASIC consists of three combined pixel matrices (Fig. \ref{fig:example}), each handling 256x256 pixels with the individual analog read-out channels. The upgraded VELO will consist of 208 such assemblies, mounted on 26 stations along the parallel axis to the proton beams, similar to the old VELO structure. Thus, the total number of channels in the detector will amount to 42 million, which is 200 times more than in the old VELO, where the corresponding number was just over $172000$. Besides the enhanced planar resolution, the nominal read-out frequency of the full event will be 40 MHz, with the innermost sensors being closer to the proton collision region than it was before \cite{Collaboration:1624070}.

\subsection{Real time data processing and analysis}

%%%decoding and low-level processing%%%%
For the regular data transmission, VeloPix uses the binary Gigabit Wireline Transmitter (GWT), which is a simplified version of Gigabit Bidirectional Transmitter (GBT) \cite{article_Hennessy}, an optical link prepared for the LHC detector upgrades. The GWT was adjusted to the front-end chip requirements and takes part in time ordering of the data packets from the read-out circuit. Optional transmission through the Experimental Control System (ECS) allows operating in an analog mode, which is essential, among others, at the charge collection efficiency measurements, as the standard GWT protocol does not support any analog information. The ECS can register a signal either in the count mode or the Time-over-Threshold (ToT) mode, where it assays how many clock cycles the signal lasted in the front-end electronics. The discriminator compares the signal at the output of the front-end with a signal threshold equivalent to the 1000 electrons. Since the front-end discharging component operates with a constant linear pace, the ToT is proportional to the charge generated by a charged particle within the sensor's active material, which makes the ECS output very useful in VeloPix performance studies. However, a fully analog protocol has many drawbacks, such as a low transmission bandwidth, which excludes it from work at the VELO operating conditions. Hence, the usage of ECS will be eventually limited, leaving the primary GWT read-out responsible for the standard data taking.

\begin{figure}[!]
    \centering
    \includegraphics[scale=0.48]{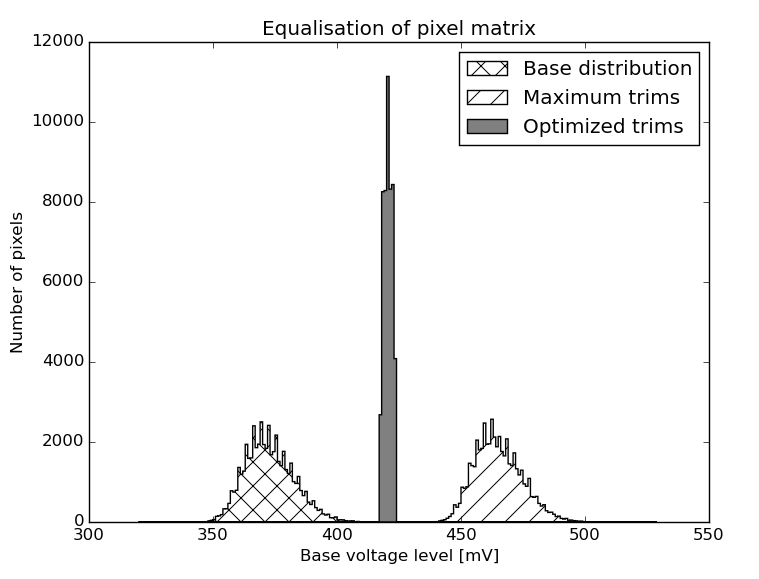}
    \caption{The outlook of pixel equalization procedure; a first scan measures the raw pedestals (distribution on the left) and makes a decision of masking if the corresponding standard deviation is unusual. A second scan applies maximal trims (distribution on the right), and basing on that, the procedure searches for the optimal trim value for every single pixel (distribution on the middle).}
    \label{equalisation}
\end{figure}

Maintaining the primary detector functionality will require regular pixel calibration. A composition of calibration scripts will create a software procedure, which must be run separately for each sensor. Such a procedure generates a recipe, which, among others, includes the calibration thresholds and the masking map for every pixel matrix. Communication through the GWT is sufficient for the whole process, although some of the more sophisticated control scans will still need analog support from the ECS transmitter path. 
During sensor read-out, the voltage signal at the output of front-end pre-amplifier is sampled by a comparator with a clock frequency of 40 MHz. The signal threshold, being a reference level for the discriminator, is set at the constant value as the equivalent of the aforementioned 1000 electrons. The static voltage at the front-end output is called a pedestal and is different for each pixel on the matrix. Keeping the detection threshold uniform requires equalization of the pedestals in the pixel read-out channels. A specific component for this purpose was introduced and installed in each channel; a collateral 4-bit local threshold trim adds to the pedestal increasing its voltage. Combining a proper set of trims on the matrix leads to the equalization of the pixel pedestals. The procedure searches for the optimal trim suite performing threshold scans, where each pixel pedestal is found as well as its fluctuation in the form of standard deviation. The outlook of the equalization procedure is shown in Fig. \ref{equalisation}. Besides equalization, the calibration framework can wield a couple of detector state checking scans. The framework is named Vetra after its predecessor in the old VELO detector and provides both real-time data processing and the offline analysis of calibration data. The new functionality of Vetra will be storing the historical data and performing a long-time trend analysis, which will be essential for the detector performance studies and radiation damage monitoring. In comparison to LHC Run 2, the expected fluence will be five times higher, and such studies will be more important than before. Historical data will be stored in a suite of database baskets and are explained more thoroughly in Section 3. 

\section{Database and trending infrastructure}
\label{db-api}
\begin{figure}[]
    \centering
    \includegraphics[scale=0.63]{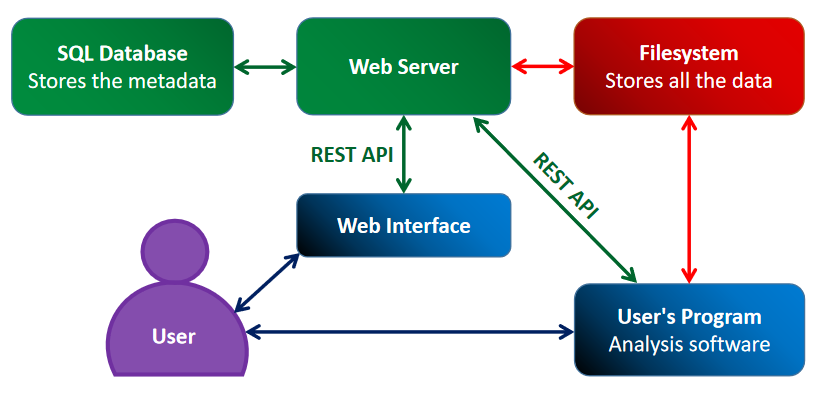}
    \caption{\label{pics:graf} Flowchart of communication within STORCK. The web interface as well as on-hand analysis software communicate with the server through the REST API protocol. A direct link between the software and the storage is also available.}
\end{figure}

The large data volume problem is common in the field of experimental high energy physics. A vast amount of data from the operating detector, even after the filtering processes or zero-suppression, are difficult to manage. 
The most straightforward solutions, like system storage or SQL database, are insufficient. Storing the data with a file system structure creates issues of access privileges and versioning. The SQL databases are known for not being efficient with large binary files and capable of a smooth data structure modification. The proposed data handling platform, based on enterprise solutions, solves these issues and delivers a better way of storing and managing data. This novel platform has been named STORCK  (\textbf{Stor}age and File Tra\textbf{ck}ing System). As seen in Fig. \ref{pics:graf}, the database system is composed of the following elements:
\begin{itemize}
  \item Web Server with User's Interface
  \item The SQL database with Metadata
  \item Filesystem Space Farm
\end{itemize}
%%%%%%%%% TO PONIŻEJ SKOPIOWAŁEM PO POPRAWKACH Z GRAMMARLY: %%%%%%%%%%%%%%%%%%%
Users can communicate with the Web Server through the Web Interface, which is a visual support for the managing process. While the SQL database stores the file paths and metadata, the actual data storage is located in the file system space. 
Each user has to be registered by a system administrator. Each authorized user can interact with the stored data either via a Web Interface or a custom User's Program. The Web Interface implements REST API in JavaScript and creates a graphical GUI that allows browsing, downloading, and uploading the data. It also enables the management of access privileges for other users and supports creating programs that can interact with the STORCK. The significant advantage of using REST API is the flexibility of communication. The user is allowed to communicate with the service in a programmatic way, using the specific access tokens.

Python is the high-level programming language of choice for many LHCb projects, consistently gaining popularity in the scientific community. Therefore, it has been decided to employ Python's Django library for creating such a database service. Users can download and upload the data using HTTP.  They can also ask the service for direct access to the file system, bypassing the bottleneck of HTTP requests when downloading larger chunks of data. The database stores paths and metadata but does not store the content of the files. Instead, files are stored in the file system, while the necessary paths of locations are stored in the database. The file system provides read access to the data but does not allow modification access for anyone except the system program (the server).
The novelty of this solution treats the database as a service, not only as static disk space. User's account restricts access to the service but every user can create his own workspace - an equivalent of a disk. The user can then upload and download any file to his workspace for the sake of simplicity and clarity. Data are not limited in term of downloading restraints. 
Files in the workspace and filesystem can create complex tree structures, with no imposed composition of such structure. The workspace administers access by sharing a special unique identifier token. Since an SQL database stores only metadata, it makes the handling of duplicates and version control possible. Files uploaded to the same path will not replace its original version but create a new record in the database, that will be considered the latest version of the file with the pointer to the previous version. When the service receives the file, its hash is calculated and compared with other hashes in the database in order to detect duplicates. If the exact hash is found, a pointer is created to refer to the existing file.

\section{Monitoring}

Continuous detector monitoring was already a critical element in software infrastructure in the VELO detector before the upgrade, when the special software monitoring platform Lovell \cite{autom_2017}\cite{Majewski_2019} was introduced. From the detector point of view, one can distinguish several aspects of monitoring. First of them is the regular data control. The complex vertex reconstruction system requires special procedures controlling the data quality; when the observed quality is compromised, a thorough analysis must be performed to verify if the problem is related to the calibration parameters or hardware components. The second aspect is the observation of the state of each sensor in the detector. If the noise inside one pixel is fluctuating too much, it may yield false hits at the read-out stage. In other words, monitoring should be able to encounter single-pixel problems and deal with them by masking or changing its local threshold trim. A generalized monitoring system will result in an independent platform of the monitoring framework.
\begin{figure}[]
    \centering
    \includegraphics[width=0.9\linewidth]{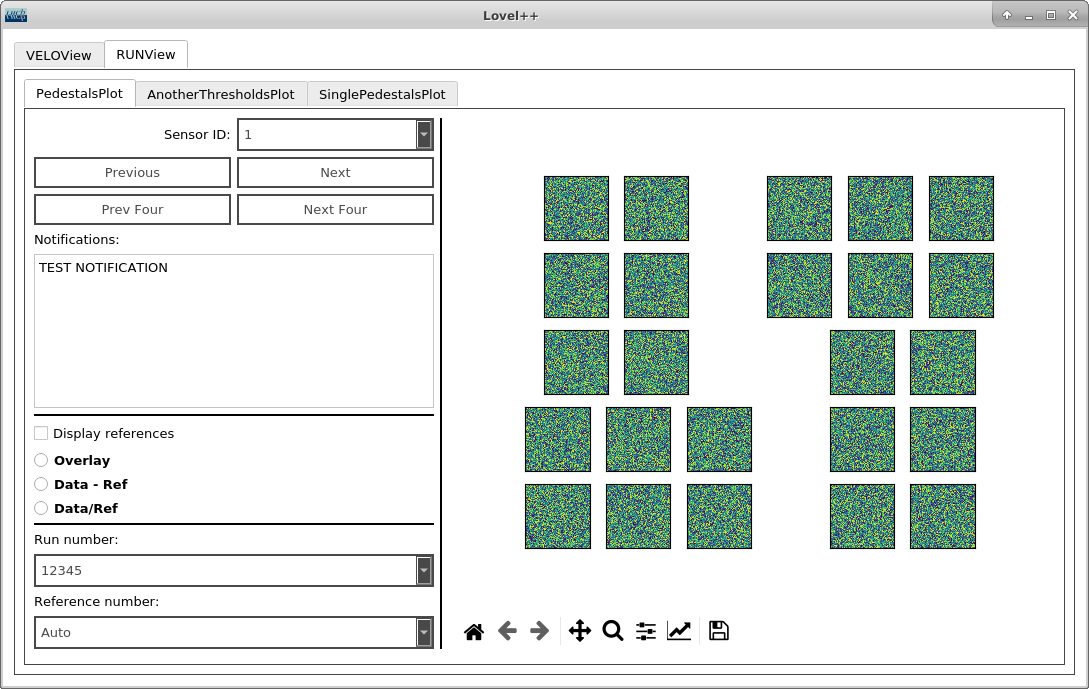}
    \caption{\label{pics:screenshot} A new version of GUI panel that is part of the new monitoring software platform. The exemplary noise map of two VeloPix parallel assemblies is presented. There is a control panel on the left side that allows switching the view between consecutive VELO sensors.}
\end{figure}
A limited number of data types and sources allows the platform to be partly autonomous. Well-defined transmission paths feature the reusability of data sources. Monitoring data will be separated from the plotting GUI as they will be useful in many other purposes in their nominal binary format.
The plot abstraction layer is defined in the form of an interface, where the special drawing method explicitly draws the demanded plot. This requires a definition of classes before the actual data acquisition. The default visual interface is programmed in PyQT5, although it was intended to web viewing with Monet \cite{Adinolfi_2017} environment as well. PyQT endpoint requires implementation of the control panel interface, which facilitates selecting desired data and plot types. The last abstraction layer is the viewing panel. The user may choose which implementation of the abstractions mentioned above should be used collaterally to create a label in the GUI or the Web app. The tasks of monitoring software are outlined in Section 3.
The viewing panel will be used for navigating the calibration processes of VELO, as well as prescribing calibration recipes. The flexibility of the platform will allow manually fitting the calibration recipes and viewing the outputs of steering algorithms. This approach creates a library-framework that will reduce the required amount of workforce spent on monitoring. Worth to mention is the fact that the proposed solution is not dedicated solely to the VELO. The generalization of concepts pertains to many potential applications, not only in the field of the specific LHCb detector physics. The compartmentalization of steps might appear excessive. However, it will support orderly iterative contributions to the monitoring platform. 
The necessity of this feature has been shown by the history of the old VELO utilization. During the LHC Run 2, the actual enrichment to the software was getting unmanageable due to a lack of internal structure of the monitoring code. We hope that this tool will find use not only in VELO detector but will provide comprehensive solutions on monitoring related tasks in other projects.

\section{Computational intelligence algorithms}
The VELO calibration data stream, because of the very large number of read-out channels, will result in a significant data volume that needs to be efficiently stored and analyzed. Also, a very high hadron particle fluence is going to introduce severe modifications in the physical properties of the silicon sensors. This constantly changing big data environment requires the application of selected computational intelligence techniques to provide a fast, efficient, and robust analysis platform. The infrastructure that provides flexible and adaptive functionality for calibration data handling has been described in Section \ref{db-api}. In this part of the paper, we focus on possible algorithms that may be useful for creating a fully autonomous system capable of independent data quality evaluation based on the calibration data stream.

\subsection{Data reduction}
The monitoring of each individual sensor is a taxing and challenging task. Instead, data reduction algorithms based on PCA (Principal Component Analysis) or ANN (Artificial Neural-Network) deep auto-encoder can be used. Both of these algorithms were tested using calibration data for the old strip VELO detector. In the case of the new detector, this approach seems to be exceptionally well suited for the analysis of the equalization procedure output data that represent a set of thresholds (trims) to be applied for each read-out channel. After the reduction transformation, the trims would form clusters in an abstract two-dimensional space. Another algorithm can subsequently be trained to detect significant changes in the properties of the clusters and trigger the re-calibration of the detector.
The granularity of the reduction (the whole VELO, per one module, or sensor) can also be tuned depending on the final properties of the system. 

\subsection{Time-sensitive analyses}
\begin{figure}[]
    \centering
    \includegraphics[width=1.\linewidth]{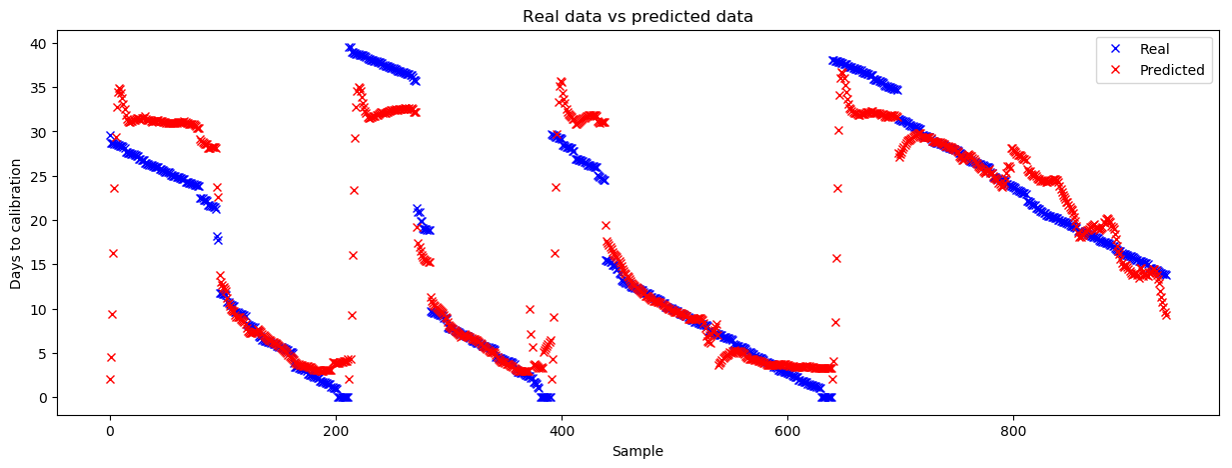}
    \caption{\label{pics:wtte} The initial results of WTTE-RNN trained to predict number of days left to next calibration \cite{Karpik:2019} (Y-axis). You can see real data (blue crosses) compared with prediction (red crosses). The samples are consecutive parameters created from the raw data coming from the detector. The performance of the model is reasonable.}
\end{figure}

Various calibration parameters or physical quantities describing properties of silicon sensors show a strong dependence on time, thus, creating respective trend plots can be a very convenient way of tracking the changes within the system. There are a number of different techniques that take into account time as one of the features of the input data. Especially interesting in the context of fully automated control software would be functionality that predicts time to next calibration based on the history of the past calibration events. A model based on the Weibull distribution WTTE RNN (Weibull Time To Event Recurrent Neural Network) has been tested for the old calibration data taken for the strip VELO detector. The results clearly showed that it is possible to build a system able to predict the time to perform the next calibration accurately. The results of the training and prediction are shown in Fig. \ref{pics:wtte}. It is worth to stress that the WTTE RNN model can take into account irregularities in the beam time (usually there are a number of random time intervals when the beam is not available) in a natural way. 

\subsection{Sensor performance anomalies}
\begin{figure}[]
    \centering
    \includegraphics[width=.9\linewidth]{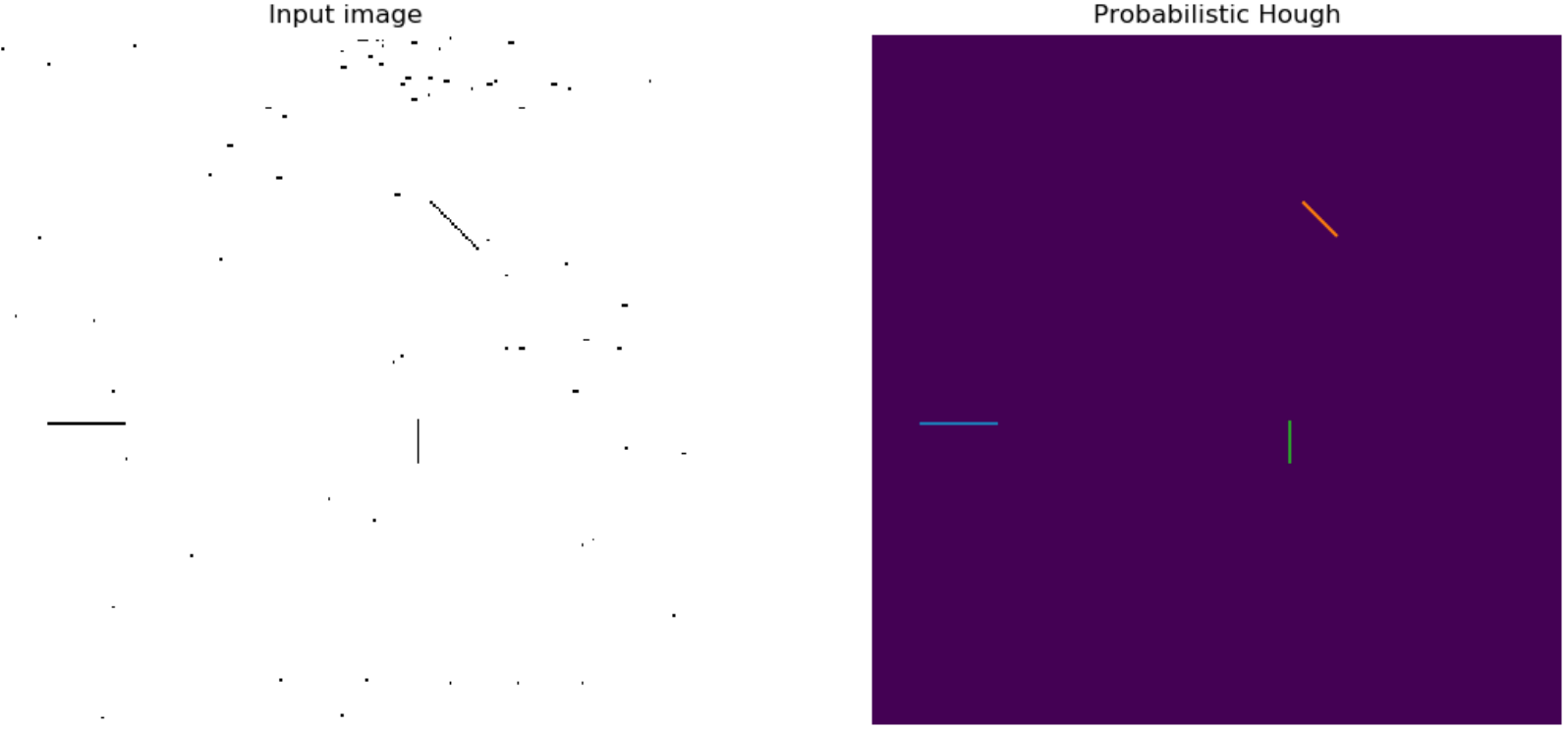}
    \caption{\label{pics:hough} The mask file reformatted as a black and white image (black point corresponds to a bad channel) with generated segments representing consecutive bad channels (left-hand side plot). The output of the Hough based algorithm with removed noise and three segments (right-hand side plot) \cite{Jarosz:2019}. The length (in pixels), start and end points are saved for the further analysis.}
\end{figure}

Fast and robust detection of anomalies during the daily operation of a complex detector system and their proper interpretation are vital and highly non-trivial tasks. One important example is related to systematic effects resulting in masking large groups of channels (e.g., consecutive pixels in a given row or column). This, in turn, may lead to unacceptable tracking efficiency losses. To mitigate this problem, we tested an algorithm, based on Hough transform, that is capable of detecting line segments on a picture. The mask files prepared for one-pixel matrix were formatted in a way to mimic a single black and white picture and constituted the input data for the algorithm. The consecutive masks were then added to the test files to check the efficiency and time performance of the code. An example of the analysis result obtained for one selected mask file is shown in Fig. \ref{pics:hough}. 

The algorithm returns as output a data structure that contains the number of pixels for each detected segment along with the start and end point coordinates. This information can be subsequently used for higher-level analysis.

\section{Summary}
The necessary change of upgraded LHCb spectrometer's running conditions during LHC Run 3 is related to the increase in instantaneous luminosity. Thus, the fast hadron fluence will increase respectively, and the tracking system must cope with much higher pile-up and track multiplicities. Also, the new triggerless data acquisition system featuring real-time tracking and event selection made the replacement of all tracking subsystems necessary. In particular, the VELO detector is going to be replaced by its modernized version. From the software point of view, a completely novel part of the new VELO, in comparison to its predecessor, will be a continuous monitoring system with the database, where the historical data describing detector states will be stored. Intelligent monitoring will lead to better performance of the VELO detector, which will eventually result in a better quality of physical data collected by the whole experiment.

\acknowledgments
We acknowledge support from CERN - Organisation Européenne pour la Recherche Nucléaire and LHCb - Large Hadron Collider beauty and from the national funding agencies MNiSW and NCN (UMO-2018/31/B/ST2/03998). We acknowledge support from National Science Centre (Poland), UMO-2017/27/N/ST2/01880.
%[@TODO add the Preludium acknowledgements]
%This is the most common positions for acknowledgements. A macro is
%available to maintain the same layout and spelling of the heading.

%\paragraph{Note added.} This is also a good position for notes added
%after the paper has been written.

\bibliography{bibliography.bib}
%\input{output.bbl}

% We suggest to always provide author, title and journal data:
% in short all the informations that clearly identify a document.

%\begin{thebibliography}{99}
%\bibitem{a}
%Author, \emph{Title}, \emph{J. Abbrev.} {\bf vol} (year) pg.

%\bibitem{b}
%Author, \emph{Title},
%arxiv:1234.5678.

%\bibitem{c}
%Author, \emph{Title},
%Publisher (year).

% Please avoid comments such as "For a review'', "For some examples",
% "and references therein" or move them in the text. In general,
% please leave only references in the bibliography and move all
% accessory text in footnotes.

% Also, please have only one work for each \bibitem.

%\end{thebibliography}
\end{document}